\documentclass[twocolumn,showpacs,preprintnumbers,amsmath,amssymb]{revtex4}
\usepackage[dvips]{graphicx}
\usepackage{color}

\def\cor#1{#1}
\def\corr#1{#1}
\def\ket#1{  \left\vert  #1   \right\rangle   }
\def\bra#1{  \left\langle  #1   \right\vert   }

\begin{document}

\title{Myoelectric Control of Artificial Limb Inspired by Quantum
Information Processing}

\author{Michael Siomau}
 \email{m.siomau@gmail.com}
\affiliation{Physics Department, Jazan University, P.O.~Box 114,
45142 Jazan, Kingdom of Saudi Arabia}

\author{Ning Jiang}
\affiliation{Department of Neurorehabilitation Engineering,
Georg-August University Medical Center, 37075 G\"{o}ttingen,
Germany}

\date{\today}

\begin{abstract}
Precise and elegant coordination of a prosthesis across many degrees
of freedom \cor{represents a significant challenge to efficient}
rehabilitation of people with limb deficiency. Processing the
electrical neural signals, collected from the surface of the remnant
muscles of the stump, is a common way \cor{to initiate and control
the different movements available to} the artificial limb. Based on
the assumption that there are distinguishable and repeatable signal
patterns among different types of muscular activation, \cor{the
problem of the prosthesis control reduces to one of pattern
recognition}. Widely accepted classical methods for pattern
recognition, however, \cor{cannot} provide simultaneous and
proportional control of the artificial limb. \cor{Here we show that,
in principle, quantum information processing of the neural signals
allows us to overcome the above-mentioned difficulties suggesting a
very simple scheme for myoelectric control of artificial limb with
advanced functionalities}.
\end{abstract}


\maketitle

\section{\label{intro} Introduction}

Information embedded within the electrical neural pulses
\cor{controlling muscle contractions in our body} can be extracted
from surface or intramuscular myoelectric signals, which are
typically summarized into so-called electromyogram (EMG). \cor{As
the acquisition of intramuscular signals might cause} ethical
concerns, may not be appreciated by the patient and can lead to
infection due to its invasive procedure, for the last forty years,
surface EMG signals \cor{have offered the primary means of
controlling prostheses \cite{Oskoei:07}}. Currently, the commercial
prostheses utilize simple processing of the surface electromyogram,
and can provide very limited functionalities \cite{Parker:04}. To
improve the functionality of myoelectrically controlled prosthesis,
pattern classification algorithms for surface EMG have been
extensively investigated \cor{in the academic community
\cite{Jiang:12}}. It has been shown, for instance, that with
properly selected features and classifiers, one can achieve very
high classification accuracy (more than 10 classes of movements with
less than \cor{5\% classification error}) \cite{Scheme:11}.

The academic success of myoelectric control based on pattern
classification has not translated into significant commercial and
clinical impact as one would have expected. So far, none of the
commercial prostheses is using pattern classification based
controller \cor{\cite{Jiang:12}}. One of the main problems with the
pattern classification for myoelectric control is that it leads to
very unnatural (for the user) control scheme. While natural
movements are continuous and require activations of several degrees
of freedom (DOF) simultaneously and proportionally, classical
schemes for pattern recognition allow activation of only one class
that corresponds to a particular action in one decision, i.e.
sequential control. Moreover, all these classes as well as their
superpositions must be previously learned. Simultaneous activation
of two DOFs is thus recognized as a new class of action, but not as
a combination of known actions. This means higher complexity of the
classifier and consequently less robustness. In addition, this also
implies that the user must spend more time in training and learning,
resulting in higher rehabilitation cost and more frustration.

\cor{Recently, the issue of simultaneous and proportional control of
a prosthesis has been addressed.} Instead of taking \cor{classical}
pattern recognition approach, for example, multilayer perceptron
neural network was used to estimate joint force \cite{Jiang:09} and
joint angles \cite{Jiang:12a} of the actions from surface EMG. One
of the main limitations of this multilayer perceptron approach is
that it still requires the data from combined activation of multiple
DOFs in the training set. Alternatively, non-negative matrix
factorization has been used to estimate the joint force
\cite{Jiang:09} and the joint angles \cite{Rehbaum:12,Jiang:13}.
This approach only requires data from single DOF activations for
\cor{the controller} calibration, but has inferior performance in
comparison to multilayer perceptron network approach. It is worth
noting that both mentioned control schemes require solving
optimization problems during the training stage, which in turn
demand significant computational power for multiple DOF's. The
training, moreover, needs to be performed on large data sets with
typical size of $10^4 - 10^5$ samples, which need to be acquired
during time-consuming sessions. Taking into account that the
classical control schemes, such as multilayer perceptron, linear
perceptron, linear discriminant analysis, Gaussian mixture model and
hidden Markov model, do not show significant difference in
classification power \cite{Hargrove:07}, in this paper we aim to
develop a radically different approach to the problem of myoelectric
control based on quantum information processing.

During the last two decades, information encoding into the states of
quantum systems and its processing according to the laws of quantum
mechanics have been demonstrating impressive advantages over
classical information processing \cite{Nielsen:00,Georgescu:13}.
Typically these advantages are discussed in the context of
computational complexity \cite{Galindo:02} and communication
security \cite{Gisin:02}. Recently we have shown that quantum
information processing can dramatically increase the capabilities of
the simplest learning machine \cor{introducing a model of quantum
perceptron \cite{Siomau:13}}. As its classical counterpart, there
are two operational stages for the quantum perceptron: supervised
learning and new data classification. During the learning stage all
the data are formally represented through quantum states of physical
systems. The subject of the learning is a set of positive operator
valued measurements (POVM) \cite{Nielsen:00}. The set is constructed
by making superpositions of the training data in a way that each
operator \cor{is designed to detect states from one class}. This
procedure is linear and does not require solving equations or
optimizing parameters. When the learning is over, new data is
encoded into the states of the quantum systems and processed with
the POVM. Based on the results of the processing, the required
classification is achieved. \cor{Further details about the quantum
perceptron could be found in \cite{Siomau:13}. \corr{It is important
to stress that, in this work, we apply only the mathematical
formalism of quantum information processing to the problem of
myoelectric control: no actual physical systems nor advanced quantum
information processing devises, such as quantum computers
\cite{Nielsen:00} and quantum simulators \cite{Georgescu:13}, are
involved.}}

\cor{The quantum perceptron superiors its classical counterpart} in
learning capabilities \cite{Siomau:13}. In particular, it is able to
perform the classification on previously unseen classes and
recognize the superpositions of learned classes. Utilizing these
properties of the quantum perceptron, in this paper we present a
novel simple scheme for simultaneous and proportional myoelectric
control of the artificial limb. We also report the first simulation
study of the surface EMG signal classification within the presented
scheme.

\section{\label{methods} Methods}

\begin{figure}
\centering
\includegraphics[width=60mm]{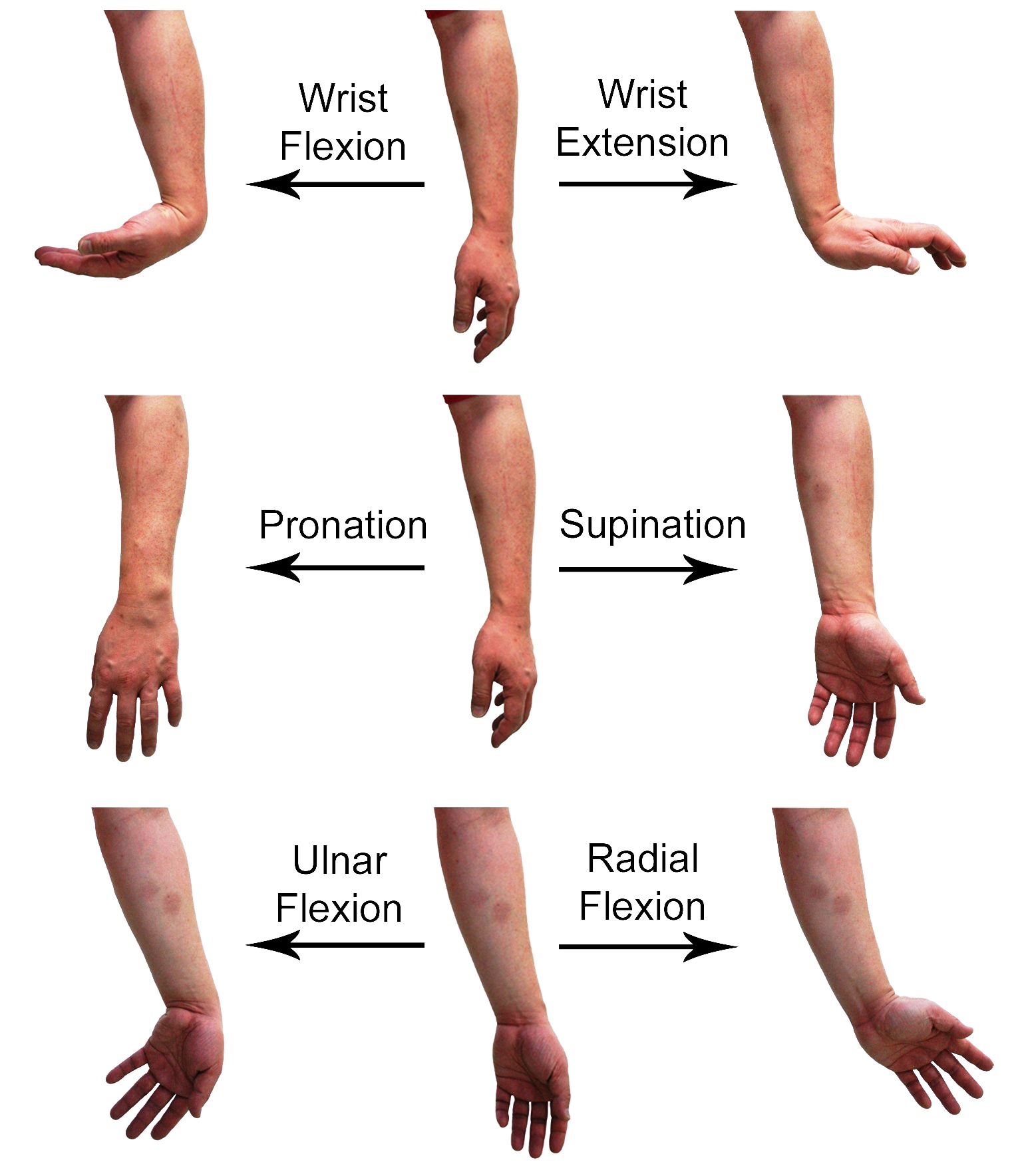}
\caption{The three DOF of the wrist to be emulated by the
prosthesis.}
 \label{fig-1}
\end{figure}

Let us focus on a particular case of limb deficiency -- transradial
amputation, or amputation at the forearm. Our choice is mainly based
on two factors. First of all, this type of limb deficiency
represents a large portion of the upper-limb amputations
\cite{Ziegler:05}. Second, wrist movements are complex and require
activation of multiple DOFs. We consider a wrist prosthesis with
three DOFs, \corr{namely flexion-extension (f-e),
pronation-supination (p-s) and radial-ulnar flection (r-u)}, as
illustrated in Figure~\ref{fig-1}.

Let us suppose that $n$ electrodes are placed on the surface of the
deficient limb. Particular choice of the location of the electrodes
strongly depends on particular amputation and remnant musculature,
therefore it is discussed elsewhere. \corr{There are four classical
features that can be extracted from the raw signals of each
electrode channel: mean absolute value (an estimate of the mean
absolute value of the signal), zero crossing (the number of times
the waveform crosses zero), slope sign change (the number of times
the slope of the waveform changes sign) and wave length (the
cumulative length of the waveform over the time segment)
\cite{Hudgins:93}.} Thus, for each of the four features the feature
space is of dimension of $n$. Let one of these features, mean
absolute value for example, be encoded into the states of a
(discrete) $n$-dimensional quantum system, so that the component
$a_i \ket{i}$ of the state \corr{$\ket{\psi} = \sum_{i=1}^n a_i
\ket{i}$ represents the feature extracted from the $i$th electrode.}
Here, the amplitude of the signal $a_i$ is normalized over
amplitudes from all the channels by factor $\sqrt{\sum |a_i|^2}$ to
fulfill the normalization condition $ \sum |a_i|^2 = 1$. \cor{This
condition leads to balanced processing of the feature vectors, i.e.
proportional control of the prosthesis.}

\begin{figure}
\centering
\includegraphics[width=80mm]{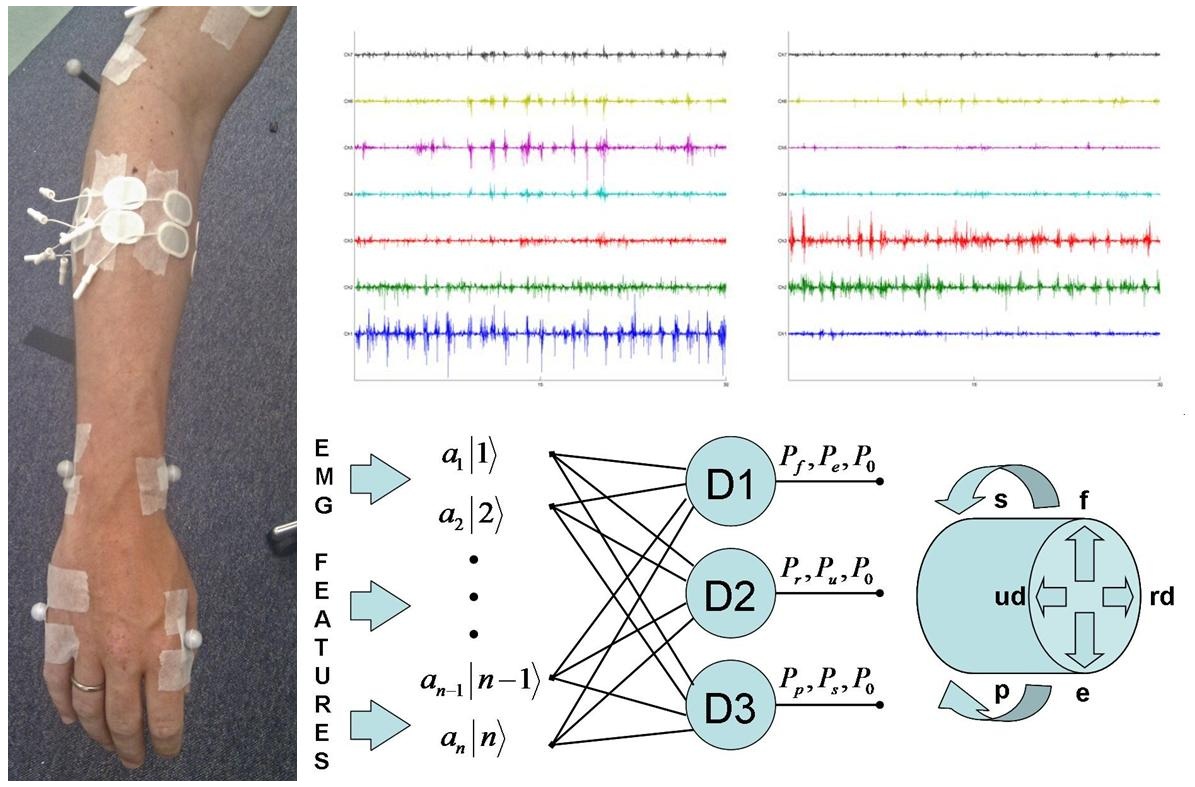}
\caption{The principal scheme of the myoelectric control of the
artificial limb. The electrodes placed on the surface of the
deficient or normal (as exemplified in the left picture) limb detect
the electrical neural signals, which are summarized in EMG (two
particular examples of EMG compound of seven electrode channels are
displayed upward). These figures reproduced from \cite{Jiang:12a},
with permission from Biomed Central. EMG features are encoded into
the amplitudes of quantum states. During the learning stage, the
three sets of positive operator valued measurements (POVM) operators
are deduced, which control the three DOFs. These operators are
utilized to decide on kind of action to be performed by the
prosthesis during the autonomous operation stage.}
 \label{fig-2}
\end{figure}

Let the quantum state $\ket{\psi}$ be the input of a single-layer
network of quantum perceptrons. In the most simple architecture,
that we are going to consider here, the number of the perceptrons in
the network is equal to the number of DOFs to be controlled, so that
each quantum perceptron governs just one specific DOF. Without loss
of generality, let us suppose that the perceptron $D1$ controls
flexion-extension, $D2$ -- radial-ulnar deviation, while $D3$ --
pronation-supination. The control scheme is represented
schematically in Figure~\ref{fig-2}.

\cor{Prior to the use of the prothesis}, the subject with the limb
deficiency is instructed to focus on performing particular actions
to learn how to control the prosthesis. Technically, the aim of the
learning stage is to deduce the association between the EMG features
and the joint kinematic. For example, given a corresponding command,
the subject is trying to perform flexion (f) activating $D1$ DOF.
EMG features \corr{extracted from received myoelectric signals} are
encoded into the quantum state \corr{$\ket{f_{j=\alpha}} =
\sum_{i=1}^n a_{i \alpha} \ket{i}$, where $\alpha$ is the angle of
flection and $a_{i \alpha}$ is the normalized signal from the $i$th
electrode that correspond to this angle}. Since efficient control of
artificial limb demands prediction of a degree of chosen action
(angle or force), the learning must be repeated for different angles
of flection. The quantum states, constructed from EMG features
corresponding to different angles \corr{$\theta_j$ are combined into
POVM operator $P_f^{D1} = \sum_j \ket{\alpha_j f_j} \bra{\alpha_j
f_j }$, where $\alpha_j = \theta_j/ \sum_j \theta_j$ and index $j$
runs over all possible angles.}

\corr{In general, if we are given just two feature vectors
\cor{$\ket{a} = \sum_i^n a_i \ket{i}$ and $\ket{b} = \sum_i^n b_i
\ket{i}$} represented through quantum states of a $n$-dimensional
quantum system, and we know that the vectors belong to one class, we
can define an operator $P = \frac{\ket{a+b} \bra{a+b}}{ {\rm Tr}
\left( \ket{a+b} \bra{a+b} \right)}$\cor{\footnote{$\ket{a+b}\equiv
\ket{a} + \ket{b}$}}. This operator ensures nonzero probability of
outcome for any given linear combination of vectors $\alpha \ket{a}
+ \beta \ket{b}$, where $|\alpha|^2 +|\beta|^2 = 1$. Taking into
account that statistics of future inputs can not be deduced, it is
reasonable to accept that all superpositions of the vectors
$\ket{a}$ and $\ket{b}$ can be observed with equal probability,
\corr{i.e. the operator $P$ is symmetric with regard to vector
permutations}. This consideration generalizes straightforwardly to
the case of an arbitrary number of feature vectors.}

\corr{If two feature vectors $\ket{a} = \sum_{i=1}^n a_i \ket{i}$
and $\ket{b} = \sum_{i=1}^n b_i \ket{i}$ represent} \cor{activation
of the same action but correspond to different angles of the
activation} $\theta_a$ and $\theta_b$ respectively, the operator $P
= \ket{ \alpha_a a + \alpha_b b} \bra{\alpha_a a + \alpha_b b}$,
where $\alpha_a = \theta_a / \left(\theta_a + \theta_b\right)$ and
$\alpha_b = \theta_b / \left(\theta_a + \theta_b\right)$, takes into
account contribution of the feature vectors to the \cor{activation
of the DOF. The construction of the operator $P$ is not unique.} It
leads, in particular, to the normalization of the all action angles
to the maximal angle observed in the training data.

Following the above procedure, the learning of extension leads to
the construction of operator $P_e^{D1}$. The operators $P_f^{D1}$
and $P_e^{D1}$ summarize all training data that activate the $D1$.
However, these operators do not necessarily form a legitimate POVM.
The fundamental property of POVM operators is that they form a
complete set, therefore, we need to define the third operator
$P_0^{D1} = I - P_f^{D1} - P_e^{D1}$ to fulfill this condition. Here
$I$ is the identity operator. Operator $P_0^{D1}$ thus collects all
features that correspond to EMG activities, which do not lead to
activation of the $D1$ DOF.

\cor{In general, operators $P_f^{D1}$, $P_e^{D1}$ and $P_0^{D1}$ are
not orthogonal, therefore it could happen that the operator
$P_0^{D1}$ is negative (i.e. unphysical). Although in our
simulations $P_0^{D1}$ was always positive, the negative value of
this operator does not imply that further metamathematical
simulations are impossible. As we noticed in the Introduction, we do
not suggest using real physical systems, but applying mathematical
formalism of quantum information processing. It is also important to
note that the triple structure of the POVM set is the key feature of
the quantum perceptron that allows detecting superpositions of
different classes \cite{Siomau:13}, i.e. achieve simultaneous
control.}

Similarly to $D1$ DOF, the learning procedure is to be repeated for
the other two DOFs: $D2$ and $D3$. As the result of the learning
stage we have deduced nine operators that govern the three DOF
according to the classification of the selected feature (mean
absolute value in our case). The learning may be repeated for the
other features the same way as discussed above.

Let us now see how the trained network of quantum perceptrons
responds to the EMG during subject's autonomous control of the
artificial limb. Desiring to perform an action, the subject
generates neural signals, which are acquired from the limb surface,
expressed in EMG and encoded into a quantum state $\ket{\psi}$. Each
quantum perceptron computes the expectation values $\bra{\psi} P
\ket{\psi}$ for the nine operators. The expectation values are
interpreted as the strength of corresponding activation. For $D1$,
for example, the three expectation values are computed as $f =
\bra{\psi} P_f^{D1} \ket{\psi}$, $e = \bra{\psi} P_e^{D1}
\ket{\psi}$ and $0_{D1} = \bra{\psi} P_0^{D1} \ket{\psi}$. The
expectation values $f$ and $e$ contain information of how likely the
given EMG correspond to flexion or extension. The expectation value
$0_{D3}$ gives the probability that the flexion-extension DOF is
inactive. If $f > e$, then the prostheses makes flexion on the
degree $\frac{(f - e) \theta_f}{1 - {\rm Tr} \left( P_f^{D1}
P_e^{D1}\right)}$, where $\theta_f$ is the maximal flexion observed
in the training data and ${\rm Tr} (...)$ stands for trace operation
and gives the overlap between the operators $P_f^{D1}$ and
$P_e^{D1}$. If, in contrast, $e > f$, the prosthesis performs
extension on $\frac{(e - f) \theta_e}{1 - {\rm Tr} \left( P_f^{D1}
P_e^{D1}\right)}$, where $\theta_e$ is the maximal observed
extension. Here, we assumed that a linear combination of DOFs
corresponds to a linear combination of corresponding features. This
assumption is valid for mean absolute value, although may not be
true for an arbitrary chosen feature \cite{Jiang:09}.

As the result of the analysis of the given state $\ket{\psi}$, which
encodes mean absolute value feature, the network of three quantum
perceptrons returns three angles to the mechanical system of the
prosthesis. These angles define the simultaneous and proportional
action completely.

\section{\label{results} Results}

We have tested the proposed control scheme on a set of EMG acquired
from an able-bodied subject who performs wrist contractions. The
data acquisition procedure as well as the experimental setup have
been previously reported \cite{Jiang:12}, thus is not discussed
here. During the experiments eight electrodes were placed around the
circumference of the forearm with equal inter-electrode distance.
The number of electrodes limits the number of input feature vectors
to eight. It is also important to note that even for a single DOF
there are two types of movements: direct action from the rest
position and return action to the rest position. EMG for these
actions may differ significantly, although formally correspond to
the same spatial angles. In our analysis we always use direct
actions both for the training and recognition of new patterns.

Our training data set collects only those movements of the subject
that activate one of the three DOFs. The test data set contains
complex movements without any restrictions. The raw EMG that
correspond to all these movements were recorded with $1024$ Hz
sampling rate. Mean absolute value feature was extracted from the
EMG averaging the signals over $100$ ms window. In our analysis we
focused on just two DOFs D1 and D3, as they are the most desired
functions not yet have commercial availability from trans-radial
amputees. The purpose is to test whether our control scheme can
recognize the combinations of these movements, being trained on just
single DOF activations, i.e. the system can extrapolate
automatically from single DOF data to their arbitrary combinations.

To test the control scheme, we initially used small sets for
training consisting of just 500 data strings for each action, i.e.
flexion, extension, pronation and supination. Each data string
consists of the mean absolute values acquired from the eight
channels and the respective joint angles of the DOF activation. The
testing data contains 8216 data strings divided on 55 blocks; each
block encodes a multiple DOF activation in a certain angle range. In
15 of the 55 blocks, the classification error appeared: 5 mistakes
in flexion-extension, 9 -- in pronation-supination and once both DOF
were misclassified. To analyze the accuracy of classification, in
cases when it was successful, we used so-called performance index,
which is widely used as a global indicator of quality of the
estimation \cite{Jiang:12a}. The performance index for an $k^{\rm
th}$ DOF is given by
\begin{equation}
 \label{PI}
R^2_k = 1 - \frac{\sum \left( \widetilde{\alpha_i} - \alpha_i
\right)^2}{ \sum \left( \alpha_i - \overline{\alpha_i} \right)^2} \,
,
\end{equation}
\corr{where $\alpha_i$ is the actual joint angle of the $k^{\rm th}$
DOF, $\widetilde{\alpha_i}$ is the corresponding estimate of the
actual angle by the suggested control scheme, $\overline{\alpha_i}$
is the temporal average (i.e. the mean value) of $\alpha_i$} and the
summation is to be done over all data samples. Similarly, the global
performance of the estimator is defined through the sum over all $K$
DOFs as
\begin{equation}
 \label{PIg}
R^2 = 1 - \frac{ \sum_{k=1}^K \sum \left( \widetilde{\alpha_i} -
\alpha_i \right)^2}{ \sum_{k=1}^K \sum \left( \alpha_i -
\overline{\alpha_i} \right)^2} \, .
\end{equation}

The performance of flexion-extension recognition is found to be
$0.834$, while the performance of pronation-supination
classification is $0.224$. It is not surprising that the performance
for $D3$ is much lower than for $D1$. Previous studies also reported
lower performance on D3 than D1 \cite{Jiang:09,Jiang:12a}. This is
mainly due to the fact that muscle responsible for D3 are deep
muscles. As a results, their EMG are easily masked by the EMG of
superficial muscles, such as flexor muscles and extensor muscles.
This masking effect is particularly pronounced during combined
activations of D1 and D3. Nevertheless, the global performance of
the two degrees of freedom recognition $0.715$ is comparable to the
performance of classical schemes \cite{Jiang:09,Jiang:12a}. \corr{In
Ref.~\cite{Jiang:12a}, for example, the global performance of the
estimator based on multi-layer perceptron artificial neural network
was shown to be in the range from $0.413$ and up to $0.906$ for
simultaneous activation of two degrees of freedom. This study,
however, was performed for different groups of subjects with actual
amputations.}

\section{\label{discussion} Discussion}

In this paper we showed that quantum information processing can be
used to realize simultaneous and proportional prosthetic control
over multiple DOFs, with only training on individual DOF data.

We suggested the simplest control scheme for the wrist prosthesis,
\cor{which is free of any optimization during the learning stage and
does not demand large training data sets} (size of which is
proportional to the time \corr{spent} by a subject in a lab and by
implication his/her frustration in the procedure). We also reported
the first simulation of EMG data classification using the suggested
scheme.

Although our simulation showed promising results, the scheme for
myoelectric control by quantum information processing must be
further developed before practical utilization. Below, we consider
several ways to improve the performance of the suggested scheme.

Using more data for the training is a standard way to improve
performance of the model. We checked how the efficiency of the angle
estimation changes with the growth of the training set size. We
subsequently used large data sets for the training: 2000 samples for
each action. For testing we used the same 55 blocks of unseen data.
In result, in 10 blocks errors were detected: 3 for $D1$ and 7 for
$D3$. Moreover, the global performance increased to $0.736$. Thus,
the efficiency of the classification increased with the size to the
training set.

It is important to stress on the role of the overlap between
operators $P_f^{D1}$ and $P_e^{D1}$ and between operators $P_p^{D3}$
and $P_s^{D3}$ in estimation \cor{of the activation strength of the}
respective DOFs. Indeed, \cor{these operators by their construction}
differ specifically in those components that are crucial for
activation of the corresponding actions. It has been observed that
the overlaps vary in certain range with the growth of training data
approaching (on the large scale) some optimal value. As closer the
overlap to the optimal values as higher the performance of
estimation. In practice, closeness of the overlaps to their optimal
values would indicate that further learning is not necessary.
Finding the optimal overlap having a small training data set is an
important open problem.

Also, there is additional control information encoded into the three
expectation values $0_{D1}, 0_{D2}$ and $0_{D3}$. As we have noted
before, the expectation value $0_{D1}$ tells us the probability that
the state $\ket{\psi}$ does not represent neither flexion nor
extension. This may be interpreted as the given signal activates the
other two DOFs, i.e. $0_{D1} = D2 + D3$. Such interpretation gives
us three additional equations with three unknowns. Resolving this
system of linear algebra equations, we can deduce additional
proportionality between the activation of the three DOFs.

Finally, we would like to admit that there are many other nontrivial
ways to improve further our control scheme. For example, it may be
beneficial to include more quantum perceptrons in the network, so
that several perceptrons control one DOF (direct and return actions,
for example). It is also possible to construct a database of POVM
operators out of the training data, so that one set of operators
controls movements in a particular angle interval. A particular set
of operators can be called from the database depending on overall
intensity of the EMG. Also, different features may be encoded and
analyzed simultaneously due to the fact that quantum amplitudes are
complex numbers in general. Finally, our control scheme can be
combined with classical control schemes, since its implementation
does not demand any optimization. It is certain that further
research is required to reveal the full power of quantum information
processing in the myoelectric control of artificial limb.

\begin{acknowledgments}
This work has been supported by European Commission via the
Industrial Academia Partnerships and Pathways (IAPP) under Grant
No.~251555 (AMYO), the German Ministry for Education and Research
(BMBF) via the Bernstein Focus Neurotechnology (BFNT) Goettingen
under Grant No.~01GQ0810 and KACST grant No.~34-37. We would like to
thank Dario Farina for his comments and suggestions, Johnny LG
Vest-Nielsen for assistance in data acquisition and Ivan Vujaklija
for art work.
\end{acknowledgments}


\begin{thebibliography}{20}

\bibitem{Oskoei:07}
   \cor{Oskoei M A and Hu H 2007 \textit{Biomedical Signal Processing and
   Control} \textbf{2} 275}

\bibitem{Parker:04}
   Parker P, Englehart K and Hugdins B 2004 \textit{Control of Upper
   Limb Prosthesis} (New York: Wiley-IEEE Press)

\bibitem{Jiang:12}
   Jiang N, Dosen S, M\"{u}ller K-R and Farina D 2012 \textit{Signal Processing Magazine
   IEEE} \textbf{29} 152

\bibitem{Scheme:11}
   Scheme E and Englehart K 2011 \textit{Journal of Rehabilitation Research \&
   Development} \textbf{48} 643

\bibitem{Jiang:09}
   Jiang N, Englehart K B and Parker P A 2009 \textit{IEEE
   Transactions on Biomedical Enginering} \textbf{56} 1070

\bibitem{Jiang:12a}
   Jiang N, Vest-Nielsen J LG, Muceli S and Farina D 2012 \textit{Journal of Neuroenginerring
   \& Rehabilitation} \textbf{9} 42

\bibitem{Rehbaum:12}
   Rehbaum H, Jiang N, Paredes C, Liliana P, Ams\"{u}ss S, Graimann B and
   Farina D 2012 Real Time Simultaneous and Proportional Control of Multiple Degrees
   of Freedom from Surface EMG: Preliminary Results on Subjects with
   Limb Deficiency \textit{Engineering in Medicine and Biology Society (EMBC),
   2012 Annual International Conference of the IEEE} 1346

\bibitem{Jiang:13}
   Jiang N, Rehbaum H, Vujaklija I, Graimann B and Farina D 2013 \textit{IEEE Transactions on Neural
   Systems and Rehabilitation Engineering} \textbf{22} 501

\bibitem{Hargrove:07}
   Hargrove L J, Englehart K and Hudgins B 2007 \textit{IEEE Transactions on Biomedical Engineering}
   \textbf{54} 847

\bibitem{Nielsen:00}
   Nielsen M A and Chuang I L 2000 \textit{Quantum Computation and Quantum
   Information} (Cambridge: Cambridge University Press)

\bibitem{Georgescu:13}
   Georgescu I M, Ashhab S and Nori F 2014 \textit{Review of Modern
   Physics} \textbf{86} 153

\bibitem{Galindo:02}
   Galindo A and Martin-Delgado M A 2002 \textit{Review of Modern Physics} \textbf{74} 347

\bibitem{Gisin:02}
  Gisin N, Ribordy G, Tittel W and Zbinden H 2002 \textit{Review of Modern Physics} \textbf{74} 145

\bibitem{Siomau:13}
   Siomau M 2014 \textit{Quantum Information Processing} \textbf{13}
   1211

\bibitem{Ziegler:05}
   Ziegler-Graham K, MacKenzie E J, Ephraim P L, Travison T G and Brookmeyer R 2008
   \textit{Archives of Physical Medicine and Rehabilitation} \textbf{89} 422

\bibitem{Hudgins:93}
   Hudgins B, Parker P and Scott R N 1993 \textit{IEEE Transactions on Biomedical Engineering} \textbf{40} 82

\end{thebibliography}
\end{document}